# SEMBridge: Tagless-Final Program Semantics with Weakest-Precondition and Bounded-Checking Interpretations

**Eric Liang**
Oracle
zixuan.liang@oracle.com

**Abstract.** Formal methods provide rigorous accounts of program behavior, but practical software engineering often works through executable libraries, tests, and incremental design. This paper presents SEMBridge, a small tagless-final framework for generating weakest-precondition and bounded-checking interpretations from the same executable object programs. Instead of committing a program semantics to one abstract syntax tree and then writing separate traversals, object programs are written once against a semantic interface and interpreted into multiple meanings: readable code, concrete execution, predicate transformers, bounded counterexample search, and future proof-assistant or SMT back ends. The Python prototype implements a loop-free imperative core with assignments, conditionals, assumptions, and assertions. Across five example programs, the same tagless-final definitions generated executable state transformers and verification conditions that passed bounded checking over domains up to 729 states. The contribution is not a Scala code-generation system or a new verifier, but a compact architecture for keeping executable semantics, weakest-precondition artifacts, and bounded validation synchronized.

## 1. Introduction

Formal methods and software engineering often share goals but not daily workflows. Formal methods emphasize mathematical definitions of behavior, proof obligations, and semantic preservation. Software engineering emphasizes executable artifacts, tests, APIs, integration, and maintainability. The result is a familiar split: a semantics written in a paper or proof assistant may drift away from the library that developers actually use, while a production implementation may accumulate behavior that is never reflected in a formal model.

This paper argues that a tagless-final embedded DSL is a practical architectural pattern for reducing that split, but narrows the claim to weakest-precondition and bounded-checking artifacts. In the tagless-final style, a program is not represented first as a tagged syntax tree. Instead, the program is a host-language function parameterized by an interface of semantic operations. Different implementations of that interface interpret the same program as concrete execution, pretty-printed code, predicate transformers, bounded counterexample checks, or future verifier inputs. The engineering unit is therefore the semantic interface, not a pile of independent traversals over a tree.

The claim is modest but useful: tagless-final embedding does not replace proof assistants, SMT solvers, Redex, K, Ott, Dafny, Boogie, or Why3. Rather, it supplies an integration layer that lets software teams design small semantic cores that can feed those tools while remaining executable and familiar to programmers. The approach is especially attractive for domain-specific languages, policy languages, workflow languages, query fragments, compiler intermediate representations, and API-level specifications.

This paper contributes a framework called SEMBridge, a loop-free prototype for imperative program semantics, a weakest-precondition interpretation, a bounded-checking interpretation, and a focused comparison with richer tagless-final EDSLs that emphasize typed host-language embedding, AST reflection, and code generation.

## 2. Background and Related Work

The formal-methods side of this work starts with axiomatic and operational semantics. Hoare logic introduced assertions as a basis for reasoning about program correctness, while Dijkstra's guarded-command work developed weakest-precondition reasoning as a method for deriving and verifying programs [1], [2]. Plotkin's structural operational semantics made transition rules a standard way to define language behavior [3]. Reynolds's definitional interpreters showed how a language can be explained by an interpreter written in a metalanguage, a pattern that remains central to executable semantics [4].

The software-engineering side starts from domain-specific languages. Hudak framed embedded DSLs as a way to build languages inside a host language rather than through a separate compiler stack [5], and Mernik et al. surveyed when DSL development is justified and how it should be approached [6].

Tagless-final interpretation, as developed by Carette, Kiselyov, and Shan, offers a typed way to represent object-language programs through semantic interfaces rather than tagged syntax constructors [7]. That work demonstrated multiple interpretations, including evaluation, compilation, partial evaluation, and continuation-passing transformations. Gibbons and Wu later clarified how tagless-final embeddings relate to deep and shallow DSL designs [8]. SEMBridge applies the same underlying idea to the software-engineering problem of keeping executable behavior, weakest-precondition formulas, and bounded validation results in sync.

Puranik's recent Scala 3 tagless-final EDSL provides a broader and more implementation-rich bridge between formal methods and software engineering, with user-defined data structures, pre/postconditions, recursion, AST reflection, type reconstruction, code generation, and a blockchain-ledger-inspired use case [9]. SEMBridge is deliberately smaller and complementary: it does not target Scala code generation or rich typed data modeling, but isolates the reusable pattern by which one loop-free object program can produce concrete execution, weakest-precondition conditions, and finite-domain counterexample searches.

Existing semantics and verification tools solve related problems at different points in the design space. PLT Redex supports executable reduction semantics [10]. Ott supports rigorous semantic definitions that can be rendered for papers and proof assistants [11]. The K framework provides executable semantic definitions that can generate analysis tools [12]. Boogie, Dafny, and Why3 show the power of intermediate verification languages and auto-active verification [13]-[15]. CompCert shows the high-assurance end point of machine-checked semantic preservation [16]. Symbolic execution connects semantic reasoning to testing and counterexample generation [17]. SEMBridge is not a competitor to these systems; it is a small embedding pattern that can make the boundary between programming and formal reasoning thinner.

Adjacent data-engineering work also motivates the setting in which semantic DSLs operate. Metadata harmonization highlights the need for shared schemas and queryable representations [18]. High-cardinality categorical representation matters when semantic tools track many program entities, resource identifiers, or provenance labels [19]. Certified-radius estimation emphasizes the value of explicit uncertainty and certification calculations [20]. Automated date-format detection illustrates the recurring problem of turning messy engineering inputs into structured semantics [21].

## 3. Design Goals

SEMBridge is designed around four goals. First, program definitions should be executable by ordinary developers. Second, semantic artifacts should be generated from the same definitions rather than manually synchronized. Third, new interpretations should be addable without rewriting existing object programs. Fourth, the framework should expose a path to stronger tools, such as SMT solvers or proof assistants, without requiring every early prototype to be mechanized in a theorem prover.

| Approach | Engineering strength | Formal-methods strength | Bridge risk |
| --- | --- | --- | --- |
| Deep AST EDSL | Good for transformations and serialization | Good for syntax-directed proofs | Requires traversals to stay synchronized |
| External verifier language | Clear verification target | Mature automation and solvers | Requires translation from production code |
| Tagless-final EDSL | Programs are ordinary host-language definitions | Multiple semantics from one interface | Harder to inspect when a global AST is needed |

*Table 1. Design tradeoffs for connecting executable DSLs and formal semantic artifacts.*

## 4. Tagless-Final Semantic Architecture

The central abstraction is a Symantics interface. It contains constructors for expressions, boolean connectives, assignment, sequencing, conditionals, assumptions, and assertions. An object program is a host-language function that receives an implementation of this interface and returns a command in that implementation's target domain. This definition has no exposed AST type. A pretty-printer implementation returns text. A concrete implementation returns a state transformer. A weakest-precondition implementation returns a function from postconditions to preconditions. A bounded checker evaluates the generated verification condition over finite domains.

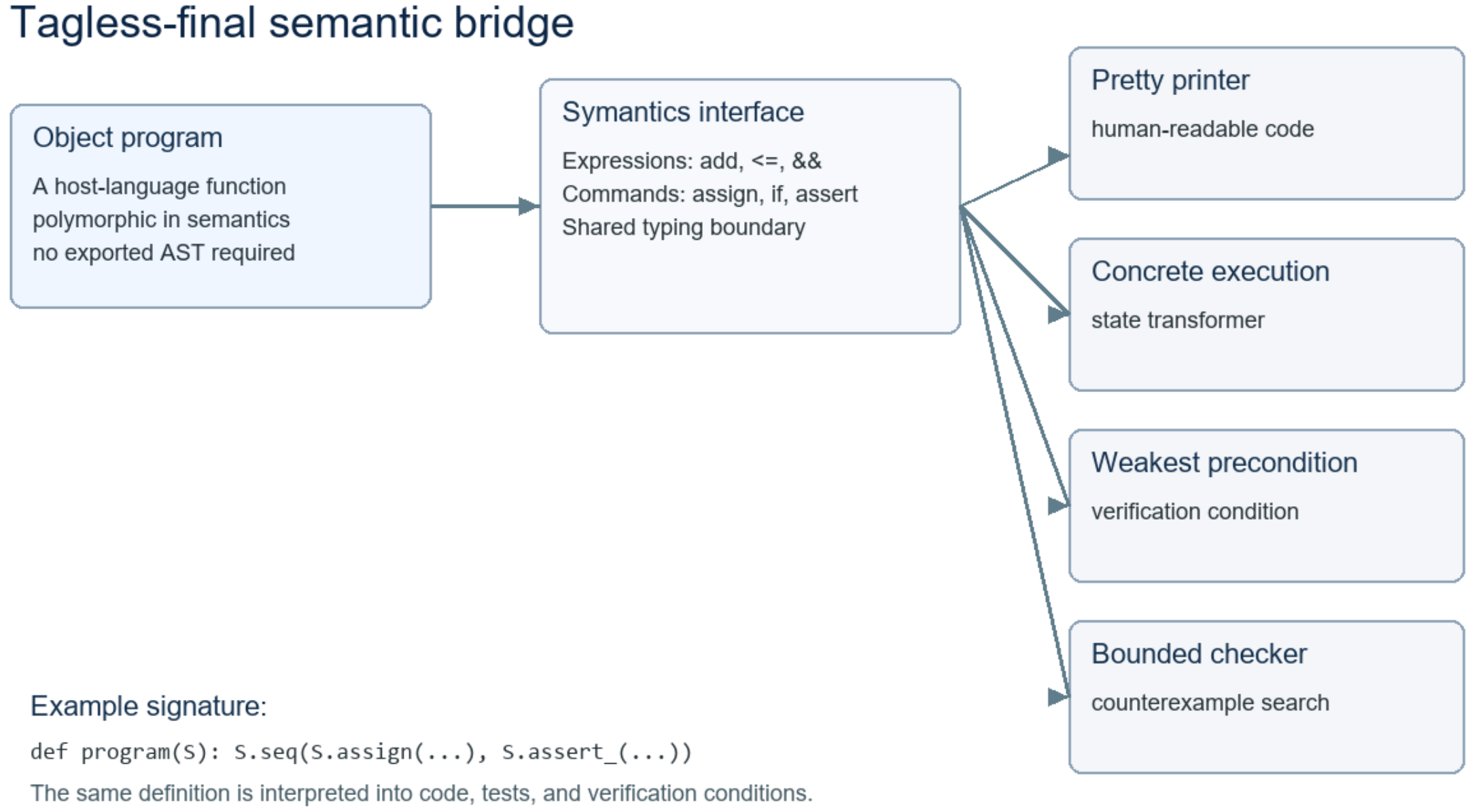

*Figure 1. SEMBridge architecture: one tagless-final object program is interpreted by multiple semantic back ends.*

The design combines shallow and deep ideas. Object programs are shallow because they call semantic operations directly. Some interpreters, such as weakest precondition generation, internally construct symbolic formulas; that is a local deep representation chosen by one interpretation, not a global representation imposed on all consumers. This hybrid is valuable in software engineering because each interpretation can choose the representation it needs without forcing every other tool to inherit it.

For the loop-free command fragment, weakest-precondition soundness follows by structural induction over command construction. Assignment substitutes the assigned expression into the postcondition. Sequencing composes predicate transformers in reverse order. Assumptions introduce implication, assertions introduce conjunction, and conditionals split verification into guarded branches. This is standard Dijkstra-style reasoning, but packaged as a reusable interpreter rather than as a separate metatheory file.

## 5. Prototype and Evaluation

The prototype is implemented in Python to keep the experiment accessible and to emphasize interpretation structure rather than host-language type-system engineering. Richer Scala 3 designs can support user-defined data structures, recursion, AST reflection, and code generation [9]; SEMBridge instead keeps the object language loop-free so that weakest-precondition and bounded-checking behavior remain inspectable. The prototype implements five example programs: absolute value, binary maximum, clamp, account transfer, and two-element sorting. Each program is written once and interpreted through four semantic back ends.

The evaluation answers three engineering questions: Can the same object programs be reused across interpretations? Are the generated verification conditions small enough to inspect? Does bounded checking find counterexamples in the intended domain? This is not a scalability benchmark and not a proof of industrial verification readiness. It is a sanity check that the embedding architecture produces coherent artifacts.

| Program | Lines | Calls | WP chars | States | Fails | WP ms | Check ms |
|---|---|---|---|---|---|---|---|
| abs | 7 | 19 | 122 | 9 | 0 | 0.075 | 0.038 |
| max2 | 7 | 19 | 135 | 81 | 0 | 0.063 | 0.289 |
| clamp | 11 | 19 | 166 | 729 | 0 | 0.063 | 2.799 |
| transfer | 5 | 23 | 135 | 729 | 0 | 0.051 | 1.313 |
| sort2 | 9 | 21 | 107 | 81 | 0 | 0.061 | 0.195 |

*Table 2. Prototype results for five tagless-final object programs over bounded finite domains.*

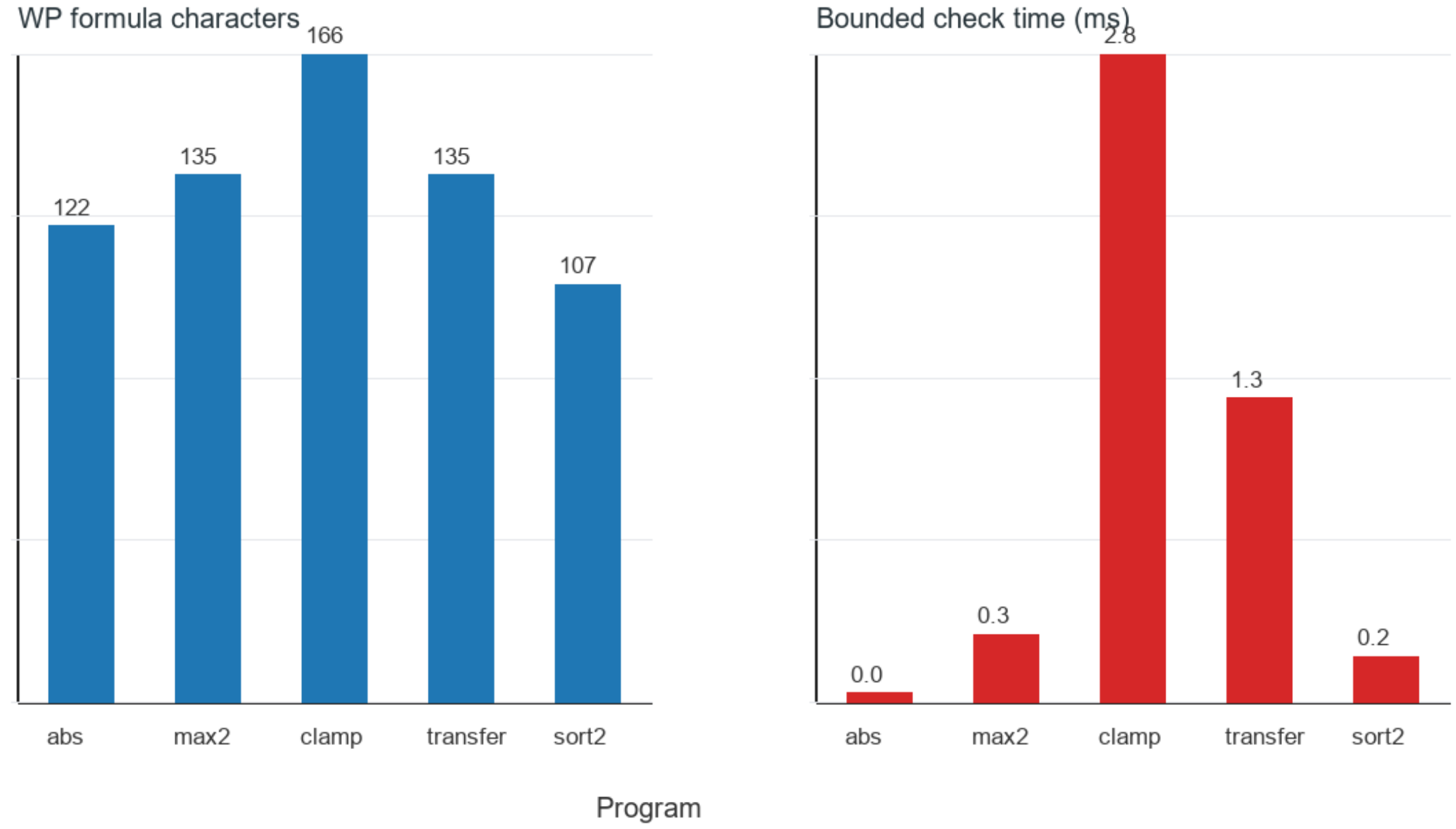

*Figure 2. Verification-condition size and bounded checking cost for the SEMBridge prototype.*

All five examples produced zero bounded counterexamples over the tested finite domains. The largest bounded search considered 729 states, and the longest generated verification condition was 166 characters after simplification. These numbers are small because the language fragment is small. The useful result is architectural: the programs, pretty-printed code, executable semantics, verification conditions, and bounded checks all originate from the same definitions.

## 6. Discussion

A tagless-final semantic DSL is useful when a team needs a single semantic source to serve multiple engineering purposes. A product DSL may need documentation, runtime interpretation, static checks, test generation, and later an SMT encoding. A compiler intermediate representation may need evaluation, pretty-printing, optimization, and proof obligations. A policy language may need executable enforcement and audit-friendly logic. In each case, the tagless-final style makes semantic interpretations explicit extension points.

The approach also changes the collaboration boundary between formal-methods specialists and software engineers. Formal-methods experts can implement interpreters that generate verification conditions or proof-assistant terms. Software engineers can write and review object programs as ordinary host-language code. The shared contract is the Symantics interface, which is small enough to document and test.

There are tradeoffs. Tagless-final definitions can be harder to inspect than explicit syntax trees, especially when programs are dynamically assembled or need source-level transformations. Some analyses genuinely require a global AST, control-flow graph, or proof term. In those cases, a tagless-final interpreter can still generate a deep representation as one interpretation. The point is not to abolish trees, but to avoid making one tree the only semantic artifact.

## 7. Limitations and Future Work

The prototype is intentionally small. It lacks loops, heap-manipulating commands, procedures, aliasing, exceptions, concurrency, quantifiers, and SMT integration. The bounded checker is not a theorem prover; it is a counterexample search over finite domains. The weakest-precondition interpreter handles a loop-free fragment only. These limitations keep the prototype auditable but prevent claims about full programming languages.

Future work should add loop invariants, procedure contracts, SMT-LIB export, source locations, proof-object export, and typed host-language implementations. Another useful direction is to combine SEMBridge's weakest-precondition and bounded-checking interpretations with the richer AST reflection and code-generation capabilities demonstrated in Scala-based tagless-final EDSL work [9]. A dual export path, where one interpreter generates executable production code and another generates Why3, Dafny, Boogie, K, or proof-assistant artifacts, would turn the architecture from a prototype into a practical bridge between lightweight engineering and heavyweight assurance.

## 8. Conclusion

Formal methods become easier to adopt when semantic artifacts are not separate from engineering artifacts. SEMBridge demonstrates a focused tagless-final pattern on a small imperative core, generating pretty-printed programs, concrete execution, weakest preconditions, and bounded checks from the same object definitions. The broader lesson is that the boundary between formal methods and software engineering can be made narrower by treating semantics as an executable, extensible interface while making each interpretation's proof and validation role explicit.

## Data and Code Availability

No external corpora or private data were used in this work. The prototype code and generated results are available from the author upon reasonable request.